\documentclass{article}

\PassOptionsToPackage{numbers, compress}{natbib}

\usepackage[final]{neurips_2024}

\usepackage[utf8]{inputenc} 
\usepackage[T1]{fontenc}    
\usepackage{hyperref}       
\usepackage{url}            
\usepackage{booktabs}       
\usepackage{amsfonts}       
\usepackage{nicefrac}       
\usepackage{microtype}      
\usepackage{xcolor}         
\usepackage{graphicx}

\title{My Voice, Your Voice, Our Voice: Attitudes Towards Collective Governance of a Choral AI Dataset}

\author{%
  Jennifer Ding\\
  The Alan Turing Institute\\
  \texttt{jding@turing.ac.uk} \\
  \And
  Eva Jäger \\
  Serpentine Arts Technologies \\
  \texttt{evaj@serpentinegalleries.org} \\
  \AND
  Victoria Ivanova \\
  Serpentine Arts Technologies \\
  \texttt{victoriai@serpentinegalleries.org} \\
  \And
  Mercedes Bunz \\
  King's College London \\
  \texttt{mercedes.bunz@kcl.ac.uk} \\
}

\begin{document}

\maketitle

\begin{abstract}
    Data grows in value when joined and combined; likewise the power of voice grows in ensemble. With 15 UK choirs, we explore opportunities for bottom-up data governance of a jointly created Choral AI Dataset. Guided by a survey of chorister attitudes towards generative AI models trained using their data, we explore opportunities to create empowering governance structures that go beyond opt in and opt out. We test the development of novel mechanisms such as a Trusted Data Intermediary (TDI) to enable governance of the dataset amongst the choirs and AI developers. We hope our findings can contribute to growing efforts to advance collective data governance practices and shape a more creative, empowering future for arts communities in the generative AI ecosystem.
\end{abstract}

\section{Introduction \& Related Work}
\label{intro}

Current concerns about AI and creativity are often grounded in artists’ fear of losing control over their work when it becomes training data for AI models. While current technical and legal discourse on this topic concentrates on enabling individual opt in and opt out, there are other dimensions of empowerment worth exploring that may be possible through collective approaches to governance that can enable further distribution of power between contributors to AI training datasets and AI developers. We introduce the “Choral Data Trust Experiment” as a case study, in particular our work surveying the attitudes of artists who contributed to the project in order to guide the design of collective governance infrastructure for the jointly created Choral AI Dataset.

Models of data governance have been explored by generative AI initiatives~\citep{jernite2022} such as the BLOOM Large Language Model (LLM)~\citep{bloom} and StarCoder LLM~\citep{starcoder}, but at present, it is not considered for many model building efforts. As a result, data contributors often remain an unacknowledged and disempowered group in the model building pipeline. This problem compounds at the intersection of arts and AI, as the work of artists has been used to train generative AI models that reproduce their work often without their knowledge, consent, or benefit. While examples are emerging for ways to expand transparency about the inclusion of art in training datasets such as Spawning’s \href{https://spawning.ai/have-i-been-trained#content}{Have I Been Trained} and to enable creator opt out such as BigCode's \href{https://huggingface.co/spaces/bigcode/in-the-stack}{Am I in The Stack?}, governance tools that focus on individual opt in or opt out put the burden to act on the individual and do not create affordances to shape the overall model building process or output. 

This project aims to challenge this status quo in AI and build upon prior efforts to capture collective rights and preferences in governance mechanisms such as ethical charters and licences \citep{pistilli2023, bigcodecard} and legal entities such as data cooperatives, data trusts ~\citep{delacroix2019, ada2021} and Trusted Data Intermediaries ~\citep{stanfordpacs2018, cdei2021} to empower data contributors to shape the process and outcomes of generative AI projects.

\section{The Choral Data Trust Experiment}
\label{cdt_experiment}

\begin{figure}
    \centering
    \includegraphics[width=.49\linewidth]{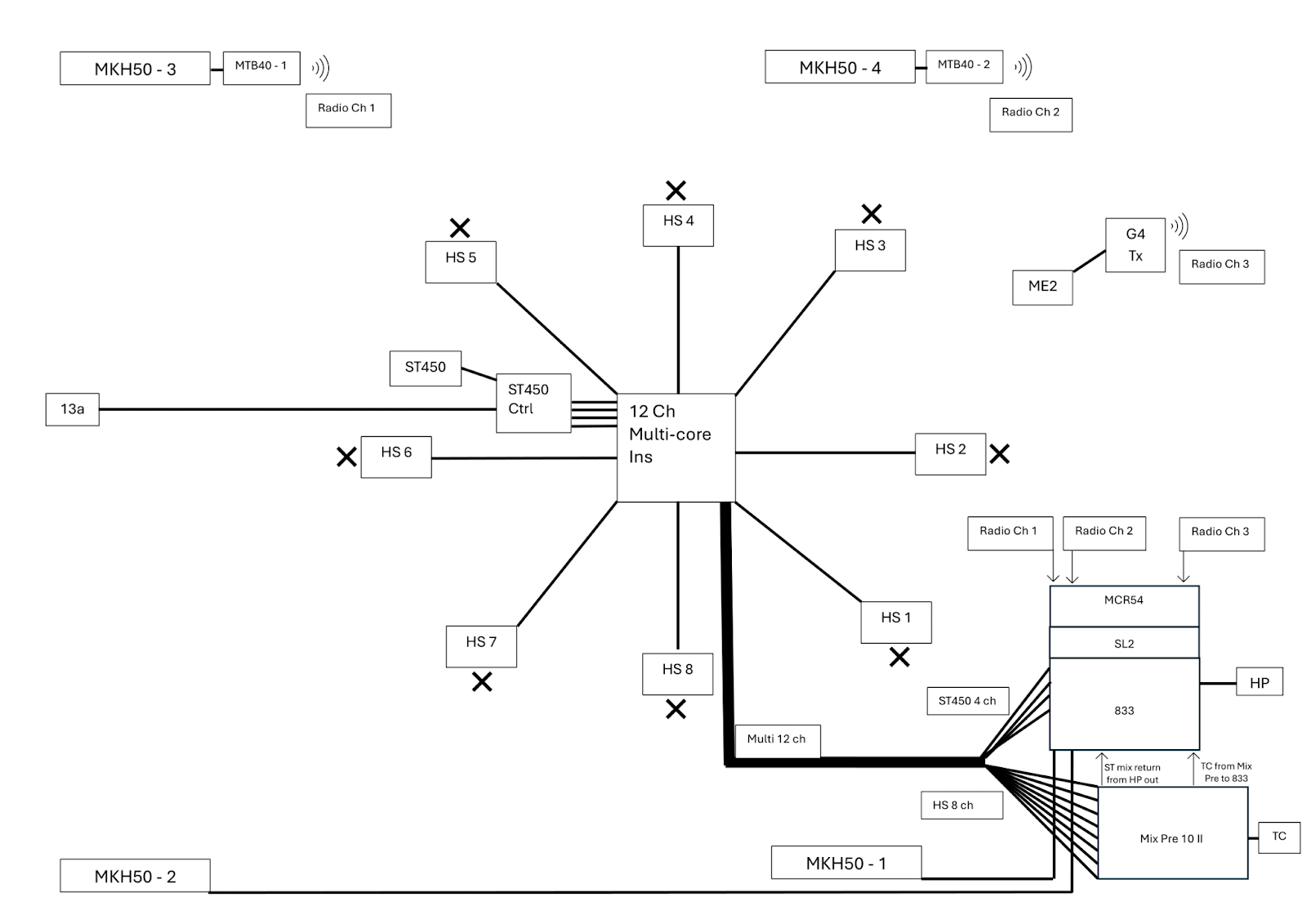}\hfill
    \includegraphics[width=.49\linewidth]{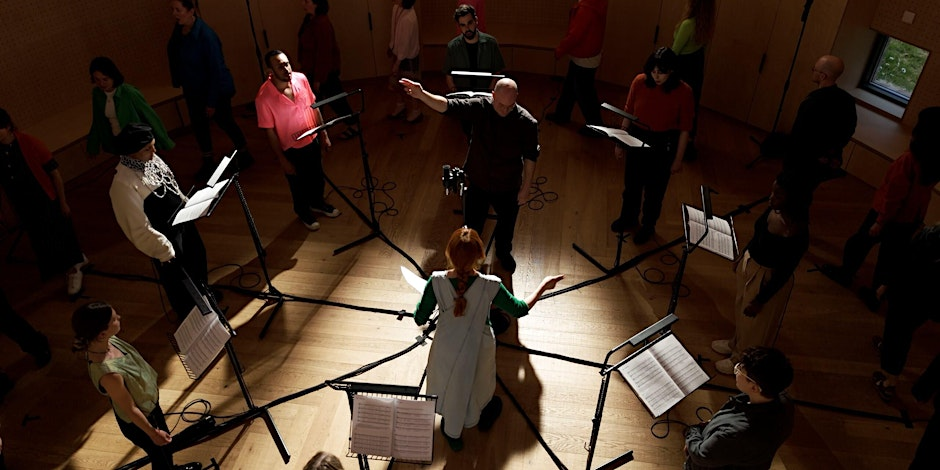}
    \caption{Schematic (left) and image (right) depicting the recording setup for collection of the Choral AI Dataset, with a multi-microphone array capturing 8 close-range microphones for soloists, 4 room microphones and a first-order ambisonic microphone}\label{fig:recording}
\end{figure}

15 community choirs from across the UK were invited to record performances of a songbook composed by artists Holly Herndon and Mat Dryhurst. The compositions and recording methods were optimised for the collection of a Choral AI Dataset, purpose-built for training Choral AI models. Herndon and Dryhurst worked alongside researchers from \href{https://www.ircam.fr/#}{IRCAM}, a French music research institute, and engineers at Stability AI to train state-of-the-art models for the exhibition \href{https://www.serpentinegalleries.org/whats-on/holly-herndon-mat-dryhurst-the-call/}{\emph{The Call}}, which opened at Serpentine in Fall 2024. This model building process is inspired by previous work by Herndon and Dryhurst on \href{https://holly.plus/}{Holly+}, a voice AI model trained using recordings of Herndon’s own voice. To collect the Choral AI Dataset, the artists traveled to each choir for the recording session, which included use of an ambisonic microphone to capture higher quality data than stereo sound to “future-proof” the dataset (see: Figure~\ref{fig:recording}). The \href{https://proud-paprika-325.notion.site/Data-Card-13debad528fe809c94b0f463d14152b5}{Data Card} for the Choral AI Dataset documents further technical information about the data collection, processing, and use considerations.

Alongside the technical challenge of scaling up data collection and model development to accommodate hundreds of different voices, this project also presents the challenge of scaling up data governance with hundreds of choristers with different backgrounds and preferences. Inspired by the \href{https://datatrusts.uk/}{Data Trusts Initiative} and the investment in human infrastructure such as Data Trustees to facilitate governance for large groups of diverse data subjects, we test the development of a Trusted Data Intermediary (TDI) to assess the opportunity for collective governance of the Choral AI Dataset.

\paragraph{Building Capacity for Collective Data Governance}
\label{build_capacity}

The TDI is composed of a team of Serpentine art curators, legal experts, and an independent data steward, who served as the primary point of contact for the choirs. The data steward began by hosting several \href{https://drive.google.com/drive/folders/1YWz-sJ-ZKqowPxuNObQcHLQzV0T9AJ49?usp=drive_link}{Data Conversations} on Zoom open to all choristers to share information about the process of training models from the Choral AI Dataset and explore points in the model building pipeline where choristers were interested in more information or agency (see: Figure~\ref{fig:pipeline_interventions}). Afterwards, a Choral Data Preferences survey was released, which received over 100 responses from each of the 15 choirs, as well as a Licence Preferences Polis, which received over 700 anonymous votes. The responses were used to \href{https://docs.google.com/document/d/1eYALwMy8zTtloRtWyJE4_Q4KVqYtj55e_BohlTsnYh0/edit?usp=sharing}{synthesise overall preferences} and \href{https://pol.is/report/r892azrash5nkrkscejbr}{identify distinct preference groups} towards use and governance of the Choral AI Dataset.

\paragraph{Data Preferences Survey Findings}
\label{survey_findings}

A key discovery was the recognition of the discomfort around using the term “data” when referring to choral performance. Choristers described this as “dehumanising”, “disembodied”, and “disempowering” while also exciting and opening up new possibilities. It also highlights the often disjointed motivations between AI developers and artists (especially live performers), where the former are focused on leveraging digital traces of the art as raw material and the latter are often focused on the ephemeral process of creating or experiencing the art itself. 

\begin{figure}
    \centering
    \includegraphics[width=.49\linewidth]{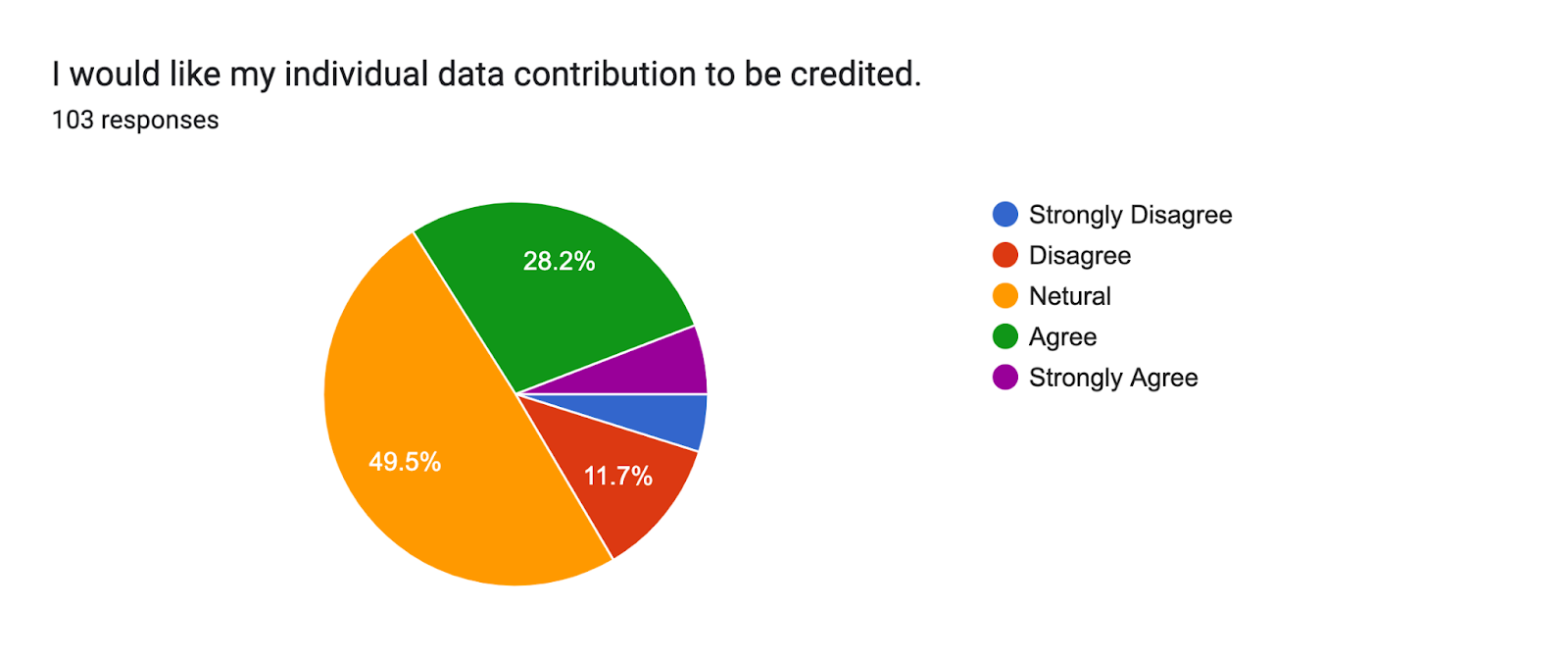}\hfill
    \includegraphics[width=.49\linewidth]{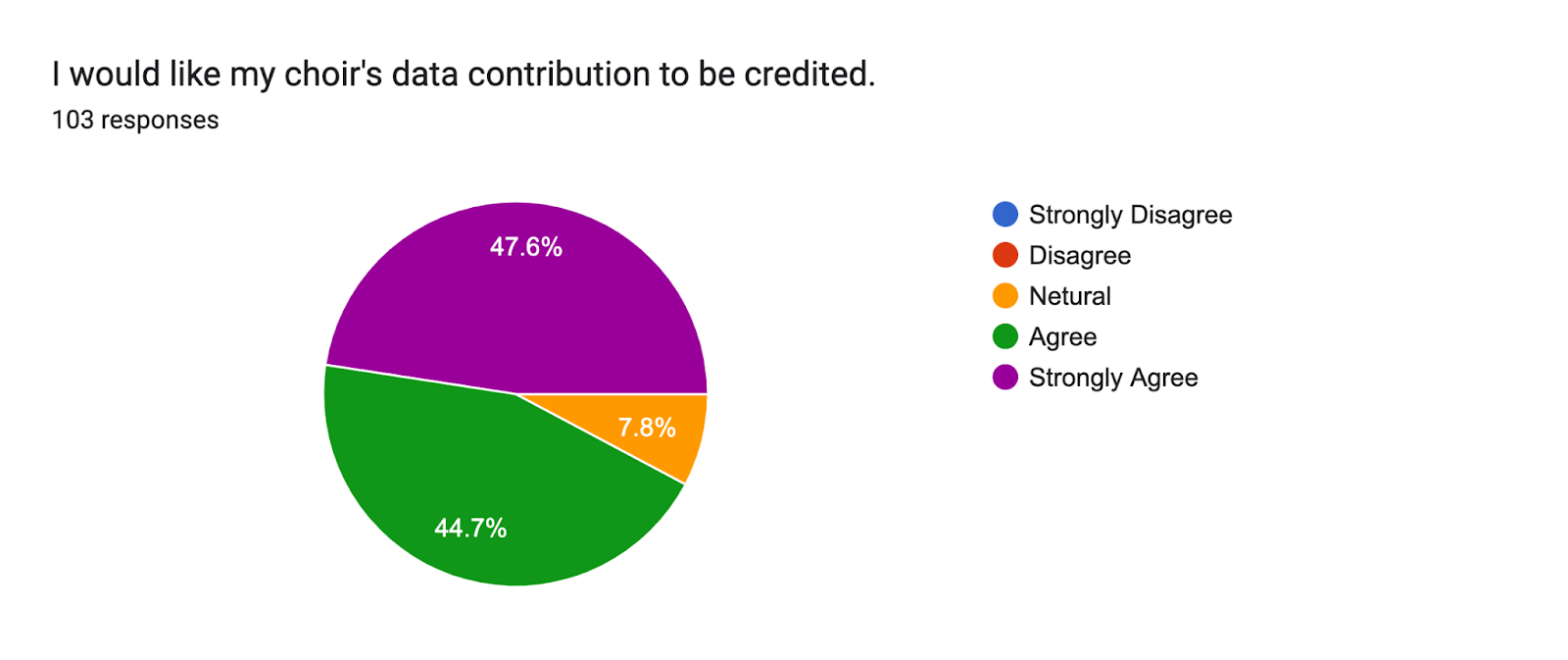}
    \caption{Changes in preferences around crediting for individual contribution (left) and choir contribution (right) by future users of the Choral AI Dataset}\label{fig:credit_chart}
\end{figure}

This concern of misalignment was reflected in survey outcomes, as over 25\% of participants responded that they were not comfortable with use of their data by users beyond the project to train an AI model. This stood in contrast to only 4\% of participants responding that they were not comfortable with Herndon and Dryhurst creating the Choral AI model for the exhibition. This indicates that transparency about data practices can mitigate unease and distrust surrounding datafication and model building processes. This also indicates that blanket opt in and opt out do not capture critical nuances of consent preferences, as context of use and user intentions may matter more than the act itself of sharing data for generative AI. The survey also highlights that participants were more interested in group recognition, with over 92\% of participants indicating interest in choir level credit, while only 34\% indicating interest in individual credit (see: Figure~\ref{fig:credit_chart}). This is further evidence that governance mechanisms should be able to interface with semantic groups rather than solely individuals.

\begin{figure}
    \centering
    \includegraphics[width=.49\linewidth]{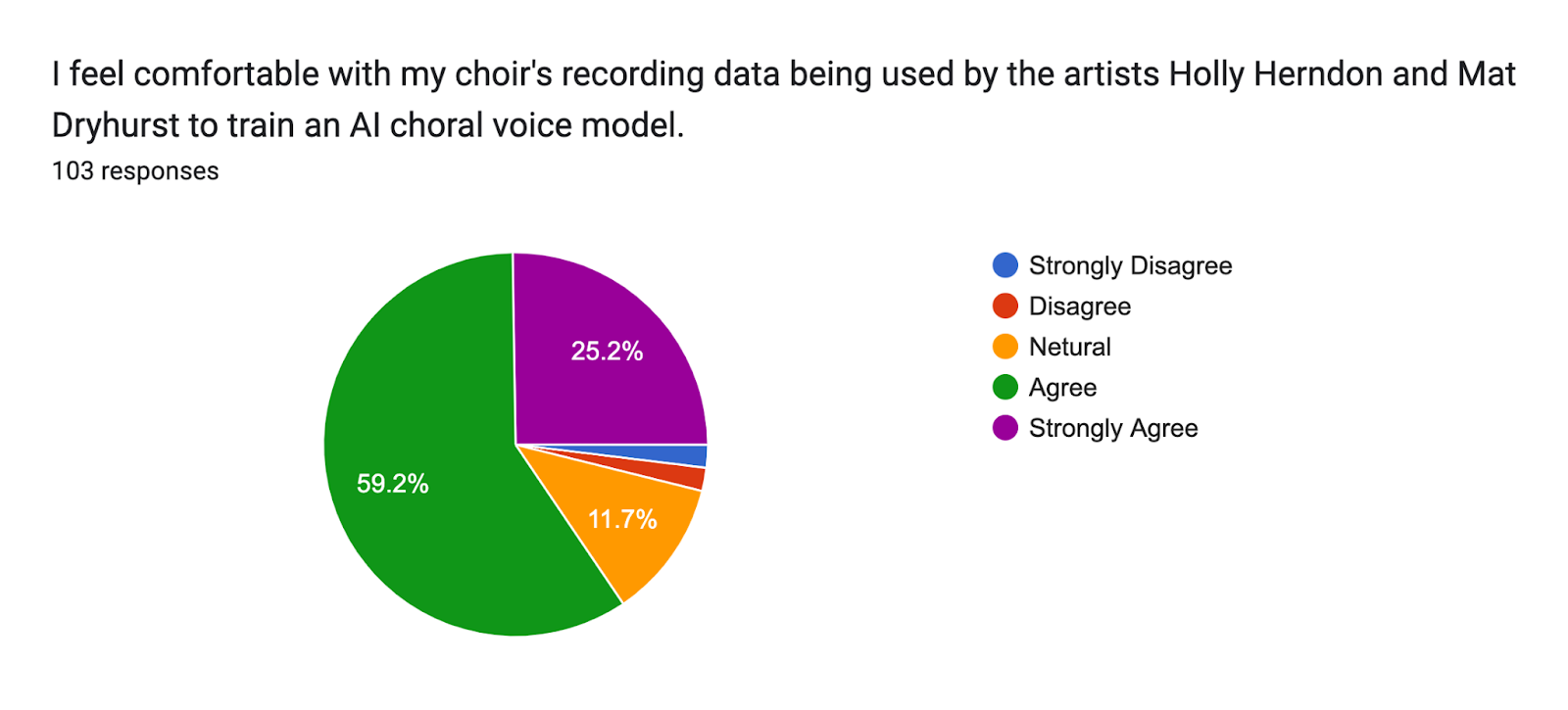}\hfill
    \includegraphics[width=.49\linewidth]{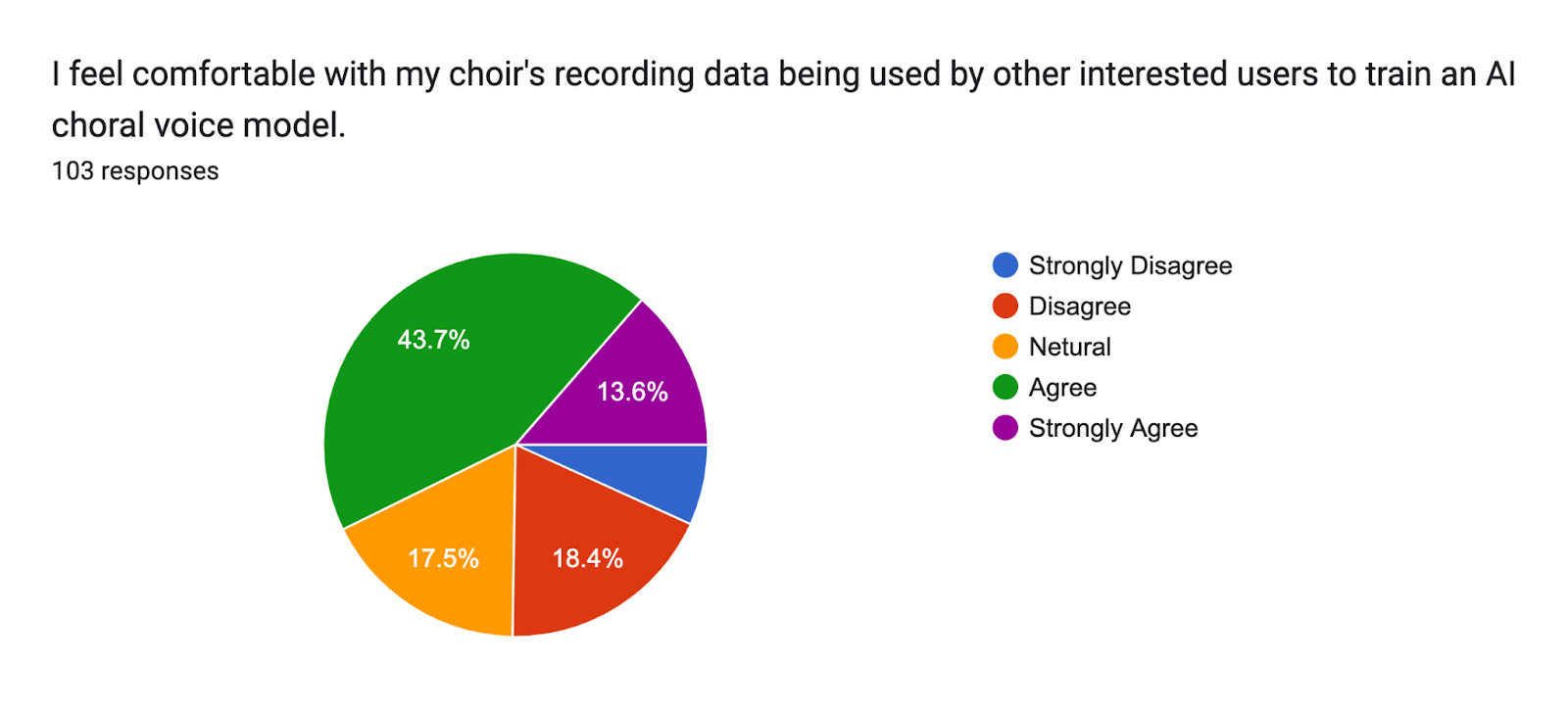}
    \caption{Changes in comfort levels around the use of the Choral AI Dataset to train models by the exhibition artists (left) and other potential users (right)}\label{fig:use_chart}
\end{figure}

\paragraph{Licence Preferences Polis Findings}
\label{polis_findings}

\begin{figure}[ht!]
    \centering
    \includegraphics[width=0.8\textwidth]{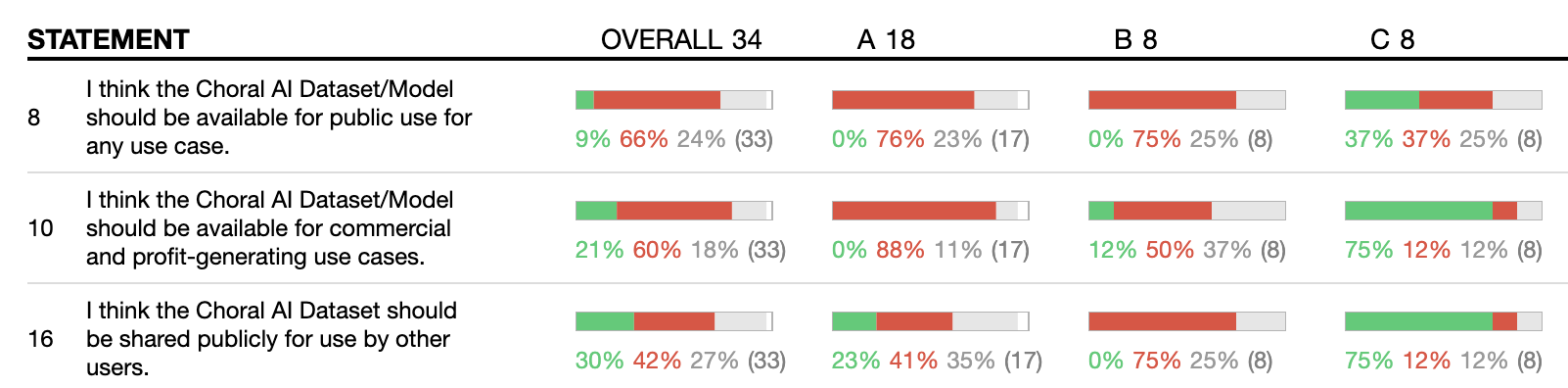}
    \caption{Polis statements with the highest levels of disagreement among preference groups}
    \label{fig:polis_disagree}
\end{figure}

Across the choirs and individuals, there were differences in risk tolerance and openness to data sharing, important considerations for setting licence terms for the Choral AI Dataset and models. To surface distinct preference groups, a Polis with 20 seed statements was shared with participants to vote on (Agree, Disagree, Pass/Unsure). These statements described different scenarios for potential dataset users and use cases. Over anonymous 700 votes were cast by 37 voters which resulted in 3 opinion groups (A, B, C). While Group C was more permissive in their views towards sharing and wider reuse of the Choral AI Dataset and models, Groups A and B were more cautious, against public sharing and commercial and profit-generating use cases (see: Figure~\ref{fig:polis_disagree}).

\begin{figure}[ht!]
    \centering
    \includegraphics[width=0.8\textwidth]{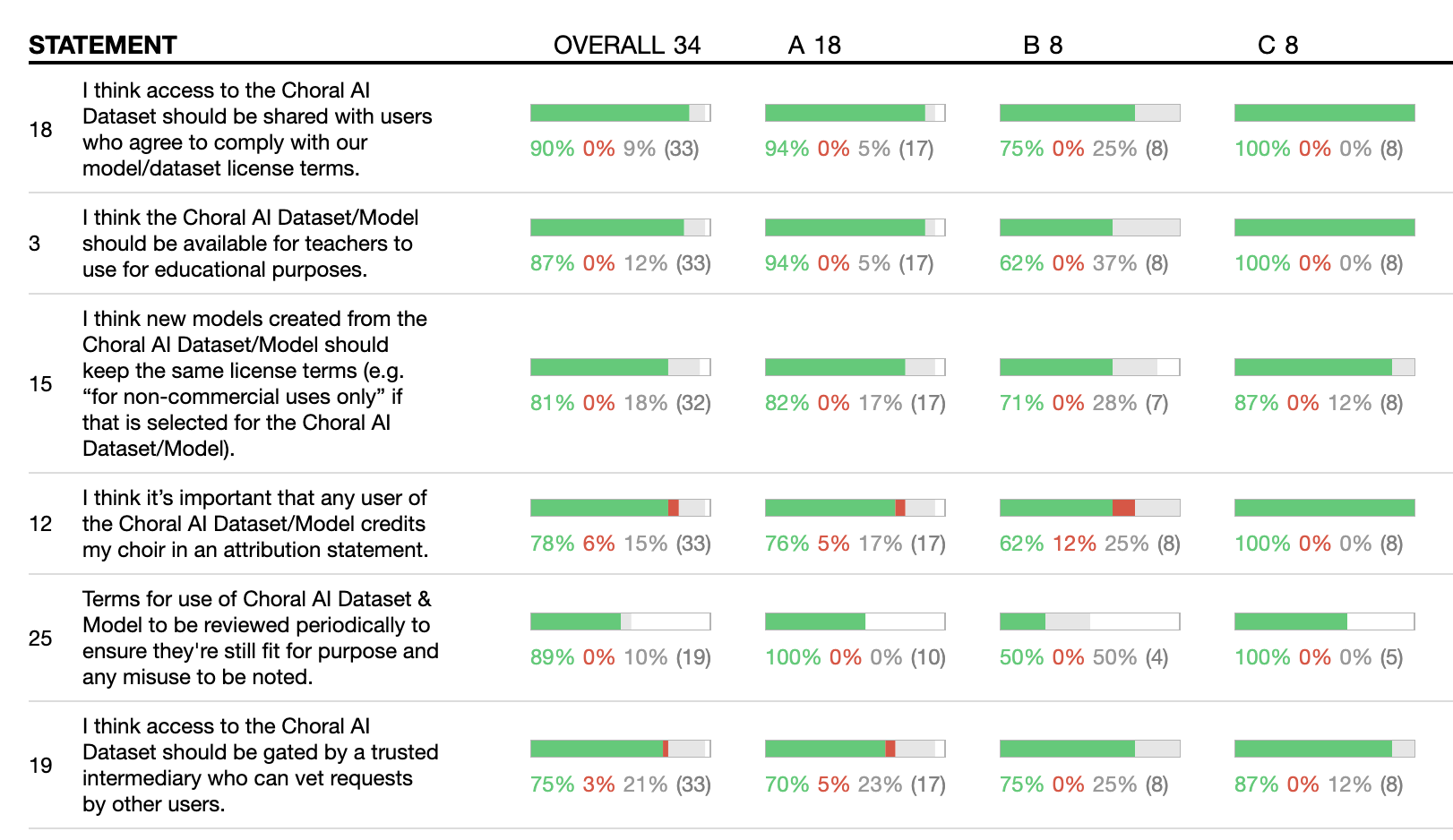}
    \caption{Polis statements with the highest levels of agreement that informed recommendations for Choral AI Dataset licence terms and further investment in the Trusted Data Intermediary}
    \label{fig:polis_agree}
\end{figure}

However, 90\% of voters agreed that the Choral AI Dataset should be shared with users who comply with the licence terms and around 80\% are interested in sharing for non-commercial use cases that re-licence under the same terms and credit the choirs for their contribution (see: Figure~\ref{fig:polis_agree}). These findings indicate that if the Choral AI Dataset is released, a licence like the \href{https://creativecommons.org/licences/by-nc-sa/4.0/deed.en}{Creative Commons Attribution Non-Commercial Share Alike} (CC-BY-NC-SA) would be a good fit to meet group preferences. This outcome and the \href{https://pol.is/report/r892azrash5nkrkscejbr}{Polis report} data will be used in future negotiations with the AI developers when selecting licences and release strategies for the Choral AI Dataset and models.

\section{Prototyping Novel Governance Mechanisms}
\label{governance_mechanisms}

Guided by the findings from the group conversations, survey and Polis, the TDI team worked with legal experts to prototype novel governance mechanisms that aimed to encode contributor preferences into actionable and accountable legal structures. These are described below and in Table ~\ref{tab:governance-table}:
\begin{itemize}
\item Formalising Serpentine LLC as the legal entity for the \textbf{Trusted Data Intermediary}, which can enter into contracts, act as the administrative hub for further sub-licensing of the dataset, and be responsible for enforcing the terms set out in the Performance Rights Agreement and Data Rights Mandate
\item Consolidating choristers’ preferences in the terms of the  \textbf{Performance Rights Agreement}, entered into between the Trusted Data Intermediary and choristers. This Agreement leverages individual  performers’ rights to set terms for downstream usage of the dataset, which includes permissible uses and types of users, expectations around data security, crediting and compensation practices
\item Creating a \textbf{Data Rights Mandate} that enables solo singers whose voices (personally identifiable information) are captured in the dataset to mandate the Trusted Data Intermediary the exercise of their GDPR UK data rights
\end{itemize}

\section{Future Work}
\label{future_work}

While the experiment is still underway, our findings raise questions about how enabling foundational components of collective data governance such as providing transparency, building trust, and accounting for diverse preferences can be managed at scale for datasets with many contributors. Alongside the development of automated tools, we propose further investigation into the development and deployment of Trusted Data Intermediaries to navigate these complex challenges. 

For the Choral Data Trust Experiment, the TDI has played an important role in capturing, synthesising, and translating preferences across the choirs into practice. By leveraging Serpentine LLC as a trusted legal entity for the TDI, we have the ability to enter into and uphold legal agreements to sustain ongoing gating, maintenance, and governance of the dataset. Whether in the form of an individual representative or team, a TDI can bring flexibility to the process of data governance and a human touch to an otherwise confusing, nonhuman process of transforming art into raw material for producing AI models. We hope to collaborate with more arts and AI communities to advance our shared understanding and best practices for collective and empowering approaches to data governance in the generative AI ecosystem.

\begin{ack}
The Choral Data Trust Experiment is led by Victoria Ivanova and Jennifer Ding, who acted as the data steward, with Eva Jäger, Ruth Waters and Mercedes Bunz. It is a research and development project associated with Holly Herndon and Mat Dryhust’s commission and solo exhibition at Serpentine, \emph{The Call} and incubated by the Future Art Ecosystems project at Serpentine Arts Technologies. 

Jennifer Ding was supported by the Ecosystem Leadership Award under the Engineering and Physical Sciences Research Council Grant [EP/X03870X/1], the Turing Institute, and Boundary Object Studio.

Further research support for this work by the Centre for Data Futures at King’s College London (Sylvie Delacroix), RadicalxChange (Matt Prewitt) and the Creative AI Lab, a collaboration between Serpentine Arts Technologies and King’s College London Digital Humanities Department.

Legal counsel for data empowerment of choir members by AWO Agency (Alex Lawrence-Archer and Ravi Naik) and Keystone Law (Alasdair Taylor).

The Choral Data Trust Experiment would not be possible without the following UK choirs:

\begin{itemize}
\item \textbf{Blackburn People’s Choir}, Blackburn
\item \textbf{Carnoustie Choir}, Carnoustie
\item \textbf{Cunninghame Choir}, Beith
\item \textbf{The Fitzhardinge Consort}, Bristol
\item \textbf{The Fourth Choir}, London
\item \textbf{HIVE Choir}, Belfast
\item \textbf{Leeds Vocal Movement}, Leeds
\item \textbf{London Contemporary Voices}, London
\item \textbf{Musarc}, London
\item \textbf{New London Chamber Choir}, London
\item \textbf{Open Arts Community Choir}, Belfast
\item \textbf{Ordsall Acappella Singers}, Salford
\item \textbf{The Ravenswood Singers}, Newcastle
\item \textbf{South Lakes A Cappella}, Windermere
\item \textbf{Spectrum Singers}, Penarth
\end{itemize}

\end{ack}

\newpage

\bibliographystyle{chicago}
\bibliography{bibliography}

\appendix

\newpage
\section{Appendix / supplemental material} \label{appendix}

\begin{figure}[ht!]
    \centering
    \includegraphics[width=1.15\textwidth]{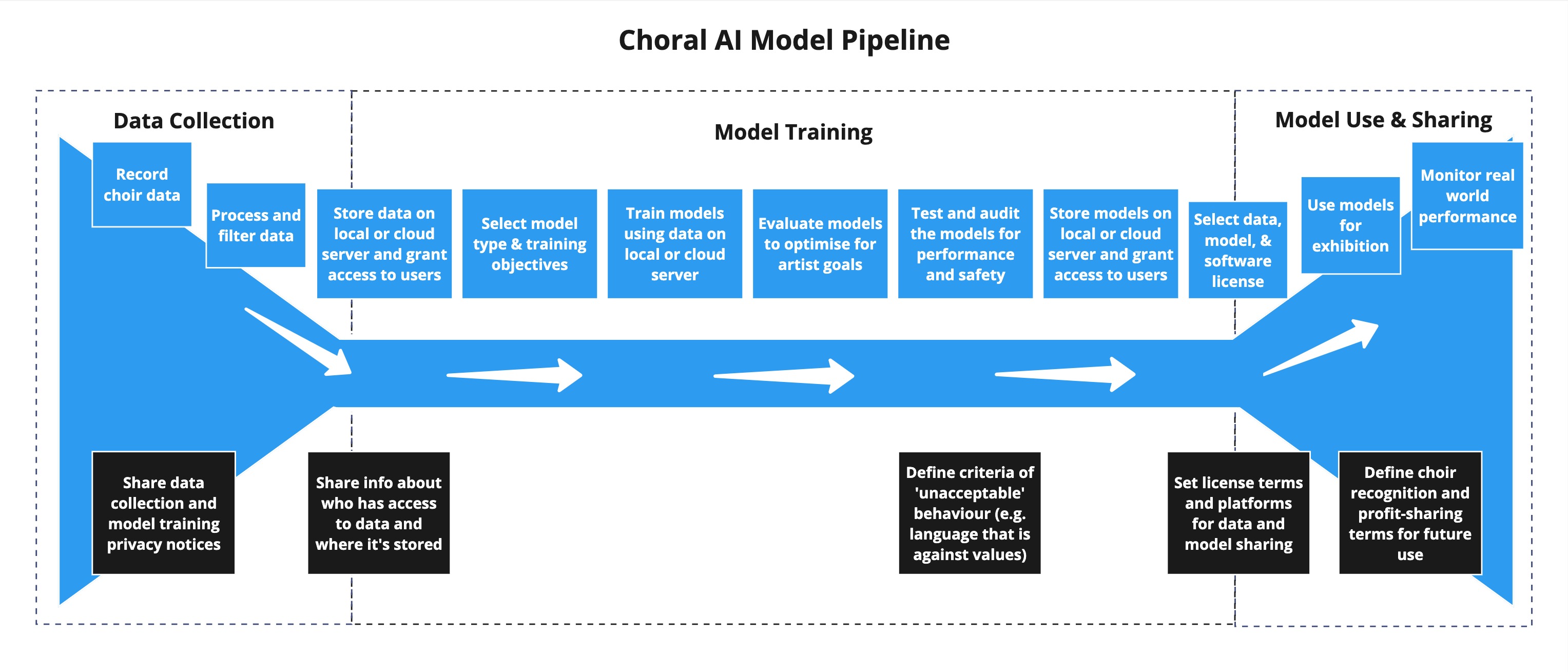}
    \caption{Priority interventions identified by contributors across the Choral AI Model Pipeline}
    \label{fig:pipeline_interventions}
\end{figure}

\begin{table}[h]
\centering
\caption{Mapping data contributor preferences to governance mechanisms}
    \begin{tabular}{|p{1.75in}|p{1.75in}|p{2.25in}|}
        \hline
        \textbf{Governance Mechanism} & \textbf{Contributor Preferences Source} & \textbf{Purpose} \\ \hline
        \texttt{Trusted Data Intermediary} &  Licence Preferences Polis & Formalises Serpentine LLC as TDI so there is a legal entity that can enter into contracts (sub-licensing), act as an admin hub and be responsible for enforcing the terms of the Performance Rights Agreement and Data Rights Mandate. \\ \hline
        \texttt{Performance Rights Agreement} & Data Preferences Survey & Set terms between TDI and choristers that collectivises individual performers’ rights for downstream usage of the dataset, which includes permissible uses and types of users, expectations around data security, crediting, and compensation practices. \\ \hline
        \texttt{Data Rights Mandate} & Licence Preferences Polis & Enable solo singers whose individual voices (personally identifiable information) are captured in the dataset to mandate TDI to exercise data rights on behalf of the group. \\ \hline
    \end{tabular}
\label{tab:governance-table}
\end{table}

\begin{figure}[ht!]
    \centering
    \includegraphics[width=0.75\textwidth]{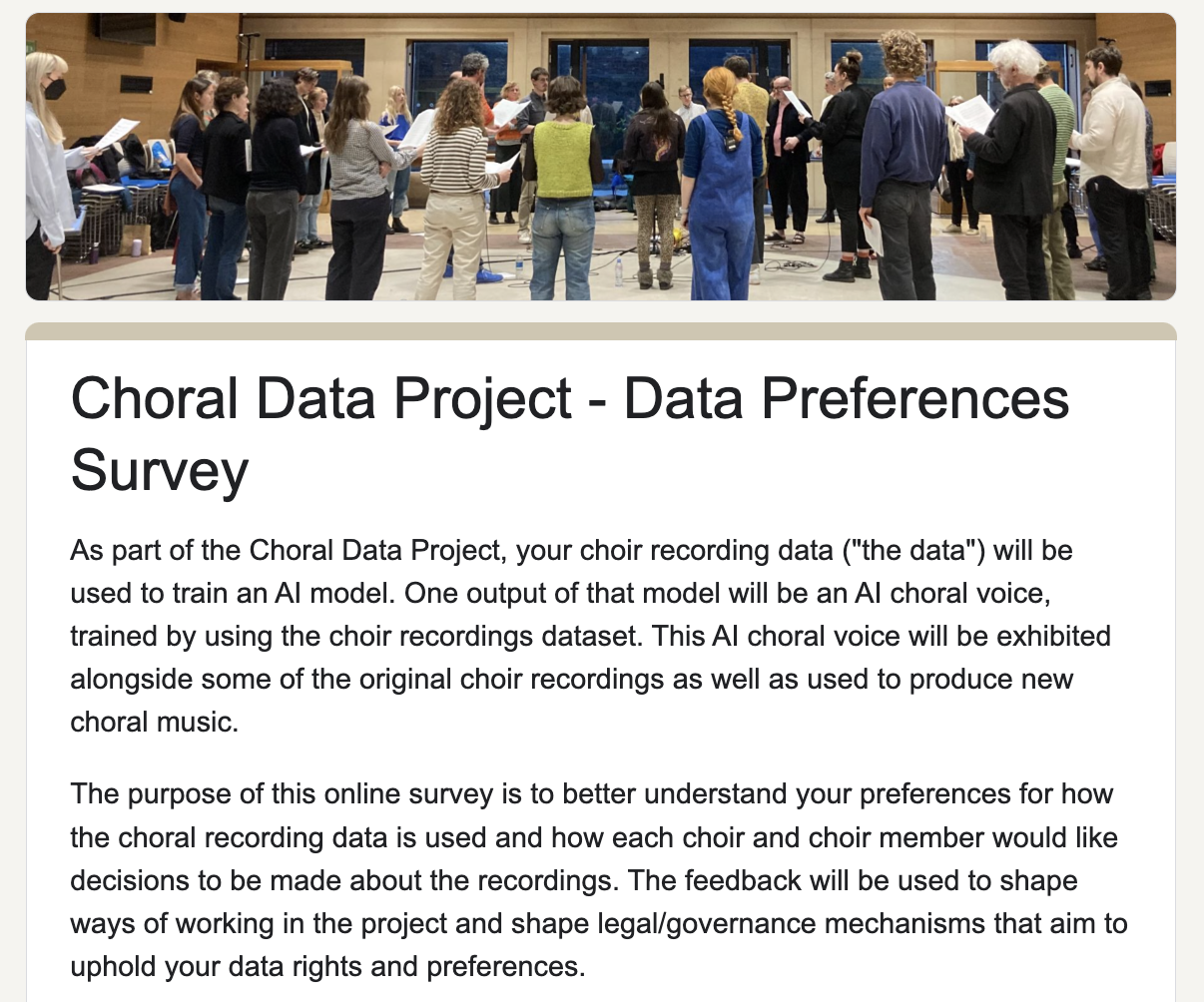}
    \caption{Screenshot of instructions for the Data Preferences Survey}
    \label{fig:survey}
\end{figure}

\begin{figure}[ht!]
    \centering
    \includegraphics[width=0.75\textwidth]{ {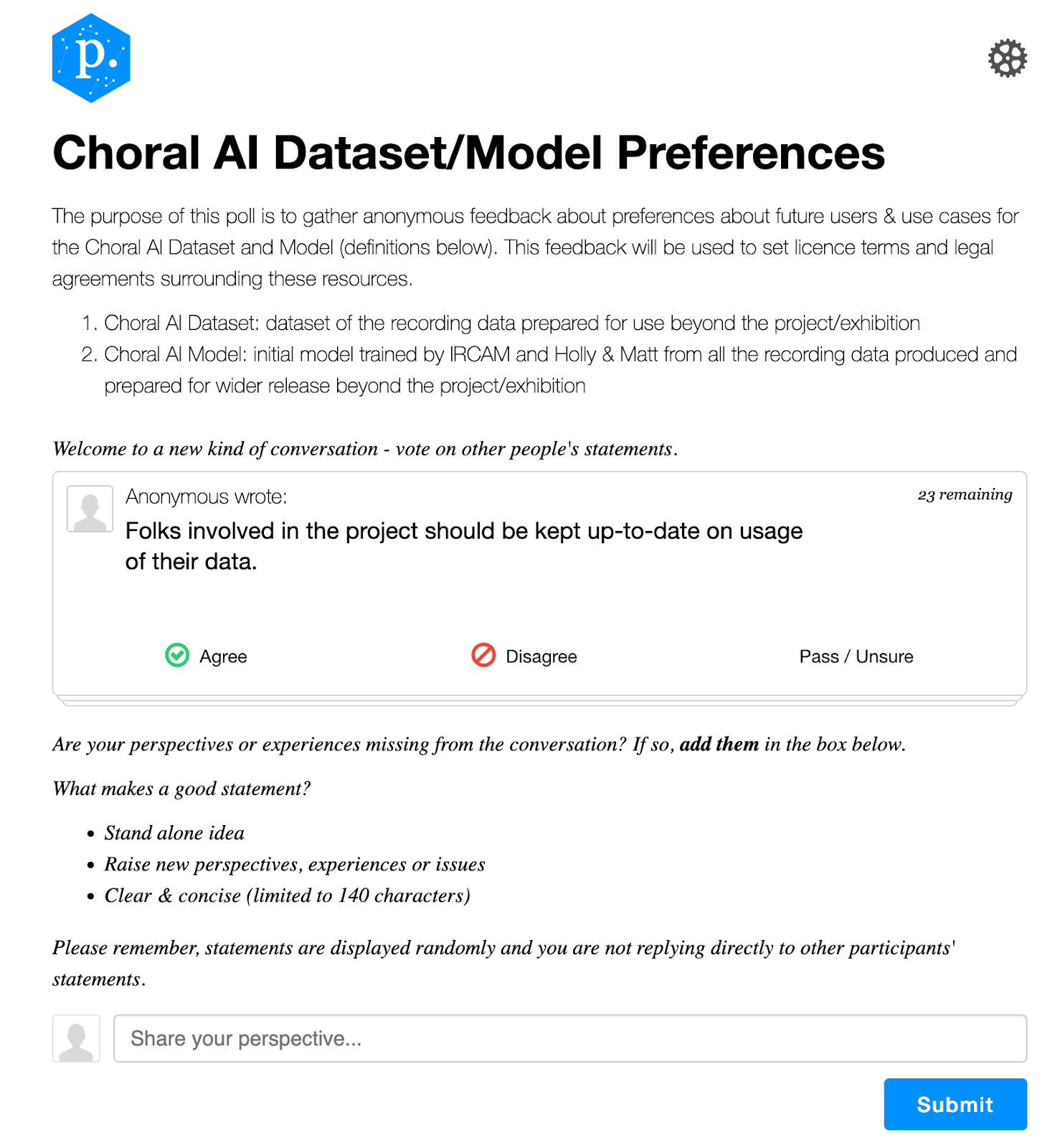} }
    \caption{Screenshot of instructions for the Licence Preferences Polis}
    \label{fig:polis}
\end{figure}

\end{document}